\documentclass[sigconf]{acmart}

\renewcommand\footnotetextcopyrightpermission[1]{}

\newenvironment{shrinkeq}[1] 
{ \bgroup
  \addtolength\abovedisplayshortskip{#1}
  \addtolength\abovedisplayskip{#1}
  \addtolength\belowdisplayshortskip{#1}
  \addtolength\belowdisplayskip{#1}
}
{\egroup\ignorespacesafterend}

\usepackage{subfigure}
\usepackage{multirow}
\usepackage{graphicx, caption, subcaption}
\usepackage{enumitem}
\usepackage{array}
\usepackage{booktabs}
\usepackage{bbding}

\usepackage{titlesec}
\titlespacing*{\subsection}{0pt}{6pt plus 2pt minus 2pt}{2pt plus 2pt minus 2pt}
\titlespacing*{\subsubsection}{0pt}{3pt plus 2pt minus 2pt}{4pt plus 2pt minus 2pt}

\AtBeginDocument{%
  }

\setcopyright{acmlicensed}
\copyrightyear{2018}
\acmYear{2018}
\acmDOI{XXXXXXX.XXXXXXX}
\acmISBN{978-1-4503-XXXX-X/2018/06}

\author{Yufei Ye}
\email{aboluo2003@mail.ustc.edu.cn}
\affiliation{%
  \institution{University of Science and Technology
of China}
  \city{Hefei}
  \country{China}
}

\author{Wei Guo}
\email{guowei67@huawei.com}
\affiliation{%
  \institution{Huawei Noah’s Ark Lab}
  \city{Shenzhen}
  \country{China}
}

\author{Hao Wang\textsuperscript{*}}
\email{wanghao3@ustc.edu.cn}
\affiliation{%
  \institution{University of Science and Technology
of China}
  \city{Hefei}
  \country{China}
}

\author{Hong Zhu}
\email{zhuhong8@huawei.com}
\affiliation{
  \institution{Consumer Business Group, Huawei}
   \city{Shenzhen}
  \country{China}
}

\author{Yuyang Ye}
\email{yeyuyang@mail.ustc.edu.cn}
\affiliation{%
  \institution{University of Science and Technology
of China}
  \city{Hefei}
  \country{China}
}

\author{Yong Liu\textsuperscript{*}}
\email{liu.yong6@huawei.com}
\affiliation{%
 \institution{Huawei Noah’s Ark Lab}
 \city{Shenzhen}
 \country{China}
}

\author{Huifeng Guo}
\email{huifeng.guo@huawei.com}
\affiliation{%
 \institution{Huawei Noah’s Ark Lab}
 \city{Shenzhen}
 \country{China}
}

\author{Ruiming Tang}

\email{tangruiming@huawei.com}
\affiliation{%
 \institution{Huawei Noah’s Ark Lab}
 \city{Shenzhen}
 \country{China}
}

\author{Defu Lian\textsuperscript{*}}
\email{liandefu@ustc.edu.cn}
\author{Enhong Chen\textsuperscript{*}}
\email{cheneh@ustc.edu.cn}
\affiliation{%
  \institution{University of Science and Technology
of China}
  \city{Hefei}
  \country{China}
}

\begin{document}

\title{FuXi-$\beta$: Towards a Lightweight and Fast Large-Scale Generative Recommendation Model}



\begin{abstract}
Recent discoveries of scaling laws in autoregressive generative recommendation models present the possibility of developing larger and more versatile recommendation systems. However, larger systems also imply increased response latency and higher training costs. To accelerate training and inference, we investigated the recent generative recommendation models HSTU and \textit{FuXi}-$\alpha$, identifying two efficiency bottlenecks: the indexing operations in relative temporal attention bias and the computation of the query-key attention map. Additionally, we observed that relative attention bias in self-attention mechanisms can also serve as attention maps. Previous works like Synthesizer have shown that alternative forms of attention maps can achieve similar performance, naturally raising the question of whether some attention maps are redundant. Through empirical experiments, we discovered that using the query-key attention map might degrade the model's performance in recommendation tasks. To address these bottlenecks, we propose a new framework applicable to Transformer-like recommendation models. On one hand, we introduce Functional Relative Attention Bias, which avoids the time-consuming operations of the original relative attention bias, thereby accelerating the process. On the other hand, we remove the query-key attention map from the original self-attention layer and design a new Attention-Free Token Mixer module. Furthermore, by applying this framework to \textit{FuXi}-$\alpha$, we introduce a new model, \textit{FuXi}-$\beta$. Experiments across multiple datasets demonstrate that \textit{FuXi}-$\beta$ outperforms previous state-of-the-art models and achieves significant acceleration compared to \textit{FuXi}-$\alpha$, while also adhering to the scaling law. Notably, \textit{FuXi}-$\beta$ shows an improvement of \textbf{27\%} to \textbf{47\%} in the NDCG@10 metric on large-scale industrial datasets compared to \textit{FuXi}-$\alpha$. 
Our code is available in a public repository: \textcolor{blue}{\url{https://github.com/USTC-StarTeam/FuXi-beta}}
\end{abstract}



\maketitle

\section{INTRODUCTION}

Recommender systems play a crucial role in modern information society. An effective recommender system not only provides users with higher quality and more personalized content in a world overloaded with information, but also creates more commercial value for enterprises. In recommender systems, increasing the model size has always faced various challenges \cite{guo2023embedding, ardalani2022understanding,xie2024breaking,yin2024learning,zhang2024unified,guo2024scaling}. However, the recently demonstrated scaling law in autoregressive generative recommendation models \cite{zhang2023scaling, zhai2024actions,shen2025optimizingsequentialrecommendationmodels,wang2025generative} offers the possibility of creating large-scale recommender systems with better performance. Although larger models improve performance, they also introduce several cost and efficiency issues: (1) Training and serving a large-scale recommendation model requires substantial GPU computational resources. 
If the computing cost significantly exceeds the commercial value it can generate, then the model loses its value. 
(2) Most recommendation applications have stringent latency constraints (usually less than 100ms) \cite{zhou2019deep,tian2023directed,zhai2024actions,tong2024mdap,wang2021decoupled,liu2023user,wu2024survey}
 , which poses significant challenges for the deployment of large-scale recommendation models.

Recent autoregressive generative recommendation models, such as HSTU~\cite{zhai2024actions} and \textit{FuXi}-$\alpha$~\cite{ye2025fuxi}, are based on the Transformer~\cite{vaswani2017attention} architecture. To address the efficiency challenges posed by increasing model sizes, we revisited recent explorations aimed at optimizing the efficiency of Transformer-like structures. Notable strategies include low-precision quantization \cite{dettmers2022gpt3, frantar2022gptq, micikevicius2017mixed}, distillation of smaller models using larger teacher models \cite{hinton2015distilling, lu2022knowledge}, and pruning of model weights \cite{han2015learning, sun2023simple, ma2023llm}. Additionally, there are numerous methods for directly designing more efficient models, such as simplifying the model \cite{he2023simplifying}, accelerating through sparse attention weights \cite{beltagy2020longformer, zaheer2020big}, utilizing kernel methods for attention acceleration \cite{choromanski2020rethinking}, employing mixture of experts models \cite{dai2024deepseekmoe}, and leveraging RNN characteristics for acceleration \cite{sun2023retentive, peng2023rwkv}. 
However, the training processes for pruning and distillation are relatively complex, as there are additional steps conducted after the base model training, and thus, they cannot directly expedite the training process. 
Moreover, quantized models necessitate hardware support, and deploying them on different GPUs may result in model failure. Some quantization methods \cite{dettmers2022gpt3, frantar2022gptq} still do not accelerate the training process. 
Although numerous efficient Transformer-like structures exist, they are tailored for NLP tasks and may not be suitable for recommendation tasks. 
Currently, the structural design for efficient generative recommendation models remains underdeveloped.

To design a more efficient structure for recommendation tasks, we analyzed the execution times of various operators during training \textit{FuXi}-$\alpha$~\cite{ye2025fuxi} and HSTU~\cite{zhai2024actions}. Among the top three operators with the highest execution times, the one with the highest execution time is associated with computing the query-key attention map. The second and third highest execution times both pertain to the indexing operator in the Relative Attention Bias (RAB) module for timestamps. On the one hand, the index operators used in RAB require noncontiguous memory access, which are not hardware-friendly, and the number of these operations is related to the square of the sequence length. These factors result in the actual runtime of the RAB being slower than the matrix multiplication, even though the latter may have a higher computational complexity. 
On the other hand, calculating the query-key attention map aims to facilitate feature interaction among items, but this process incurs significant computational and memory costs. Previous research suggests that the attention map used in self-attention does not necessarily need to be generated in this manner. For example, using a fixed attention map \cite{raganato2020fixed} or a learnable constant attention map \cite{tay2021synthesizer} can still achieve comparable results. In \textit{FuXi}-$\alpha$ and HSTU, aside from the query-key attention map, the bias term produced by RAB can itself be regarded as an attention map. This naturally raises the question: Are some of these attention maps redundant? To answer this question, we conducted empirical experiments on multiple datasets and found that removing the temporal or spatial attention maps results in varying degrees of performance degradation. However, removing the query-key attention map, conversely, improved performance. This suggests that the query-key attention map might be redundant in recommendation tasks.

To address the aforementioned issues, we propose a new framework that can be utilized in various Transformer-like models. On the one hand, we introduced the Functional Relative Attention Bias (FRAB) module, which addresses the inefficiency of RAB by avoiding various hardware-unfriendly operations inherent in RAB. On the other hand, we removed the query-key attention map from the original self-attention layer and designed a new module called Attention-Free Token Mixer (AFTM).
Subsequently, we apply this framework to the prior state-of-the-art generative recommendation model and introduce a new model, \textit{FuXi}-$\beta$.
And \textit{FuXi}-$\beta$ achieves performance comparable to the prior state-of-the-art models on public datasets while significantly reducing training time. On large-scale industrial datasets, it achieves both effective acceleration and notable performance improvements. Additionally, we have validated on industrial datasets that our model exhibits the properties of scaling laws, demonstrating its potential for application in large-scale recommendation systems. Our contributions are as follows:

\begin{itemize}[leftmargin=*,align=left]
\item We propose a new framework that can simplify Transformer-like generative recommendation models and introduce a new efficient and lightweight generative recommendation model, \textit{FuXi}-$\beta$.
\item We empirically find that the query-key attention map can have a negative impact on recommendation tasks. Additionally, we designed a new module called Attention-Free Token Mixer.
\item We propose a novel and efficient method for modeling relative temporal information, termed Functional Relative Attention Bias.
\item Experiments on four real-world datasets demonstrate that \textit{FuXi}-$\beta$ outperforms several state-of-the-art models in performance and efficiency. Furthermore, its scaling law properties suggest potential for large-scale recommendation systems.

\end{itemize}

\section{RELATED WORK}

\subsection{Positional and Temporal Embedding Methods}

Due to the inability of self-attention to capture the sequential order, it is necessary to incorporate positional encodings as an additional component. Positional encodings can be categorized into absolute and relative positional encodings. Absolute positional encodings involve directly adding a position embedding to each token's hidden representation. Common approaches include Sinusoidal positional encodings \cite{vaswani2017attention} and learnable positional encodings \cite{gehring2017convolutional}. In contrast, relative positional encodings consider the relative distances between tokens when computing attention weights. In 2018, Shaw et al. \cite{shaw2018self} introduced the concept of relative positional encodings for the first time. Subsequent works \cite{dai2019transformer, he2020deberta, ke2020rethinking, zhang2025killingbirdsstoneunifying} have proposed various forms of relative position encoding by considering the interaction terms between positions and tokens, as well as between positions themselves.
T5 \cite{raffel2020exploring} incorporates a learnable bias based solely on relative position information directly into the attention weights. Huang et al. \cite{huang2020improve} propose more sophisticated relative positional encodings, asserting that existing methods do not fully leverage the potential of relative positions.

In the domain of sequential recommendation, it is crucial to consider both positional and temporal information \cite{li2020time, zhai2024actions, han2023guesr, han2024efficient, wang2024denoising}. TiSASRec \cite{li2020time} and STAN \cite{luo2021stan} model time intervals by segmenting them into discrete bins, each associated with an embedding vector added to the key vector. MEANTIME \cite{cho2020meantime} applies various functions to the relative time differences to obtain temporal embeddings, which are then incorporated into the key vector. HSTU \cite{zhai2024actions} employs a logarithmic bucketing approach for time intervals, followed by a T5-style relative position encoding to integrate temporal information.

In our approach to modeling time intervals, we adopt the T5-style relative positional encoding but replace bucketing and learnable vector parameters with learnable functions, thereby enhancing computational efficiency.

\subsection{Efficient Transformer-based Models}

Transformer-based models have been widely applied across various domains \citep{vaswani2017attention, dosovitskiy2020image, kang2018self, tian2024visual,yin2023apgl4sr,xie2024bridging,xu2024multi,wang2019mcne}. However, the quadratic time and space complexity of self-attention with respect to sequence length imposes significant limitations. Consequently, numerous algorithmic strategies have been proposed to reduce these complexities. Techniques such as model pruning \citep{han2015learning, sun2023simple, ma2023llm}, which removes less important weights to reduce parameter count and improve computational efficiency, and model quantization \citep{dettmers2022gpt3, frantar2022gptq}, which compresses models by converting high-precision floating-point parameters to lower-precision floats or integers, are commonly used. Knowledge distillation \citep{hinton2015distilling, lu2022knowledge,zhang2025td3,yin2024dataset} further accelerates models by training a smaller student model under the guidance of a larger teacher model, effectively transferring learned knowledge.

 Additionally, designing more efficient model architectures is also an option to address these challenges \cite{hua2022transformer,sun2023retentive,he2023simplifying,wang2025mf,wang2025universal}. ALBERT \citep{lan2019albert} reduces parameters through low-rank factorization of embedding layers and cross-layer parameter sharing. Approaches such as MQA \citep{shazeer2019fast} and GQA \citep{ainslie2023gqa} decrease memory usage by reducing the number of query and value heads. MLA \citep{liu2024deepseek} builds on GQA by incorporating additional projections to enhance performance while further compressing the KV cache. However, these efforts can only accelerate the inference process. He et al. \citep{he2023simplifying} simplified the Transformer architecture by removing components such as the value projection matrix and normalization layers. FLASH \cite{hua2022transformer} achieves simplification by integrating Self-attention layers with Gated Linear Units (GLU), while reducing the complexity and memory overhead of self-attention through chunking and linear attention mechanisms. RWKVs \cite{peng2023rwkv, peng2024eagle} and RetNet \cite{sun2023retentive} modified the attention mechanism to achieve the low-cost inference characteristic of RNNs, while simultaneously maintaining the capability for parallel training.

\section{PROBLEM STATEMENT}

In the paradigm of generative recommendation, the primary objective is to predict the next item a user is likely to interact with, based on their historical interaction sequence. 
Besides the item and position information, we need to pay additional attention to the timestamp of each interaction. Formally, consider a set of users $\mathcal{U} = \{u_1, u_2, \ldots, u_{|\mathcal{U}|}\}$ and a set of items $\mathcal{I} = \{i_1, i_2, \ldots, i_{|\mathcal{I}|}\}$. For each user $u \in \mathcal{U}$, we define an interaction sequence 
\begin{shrinkeq}{-4px}
\begin{align}
\mathcal{S}_u = \left [\left (i_1^{(u)}, t_1^{(u)}, 1\right), \left (i_2^{(u)}, t_2^{(u)}, 2\right), \ldots, \left (i_{n_u}^{(u)}, t_{n_u}^{(u)}, n_u\right) \right ]    
\end{align}
\end{shrinkeq}
which is a chronologically ordered list of items, their corresponding timestamps and positions.

The task is to predict the next item $i_{n_u+1}^{(u)}$ that user $u$ will interact with, based on their interaction sequence $\mathcal{S}_u$. 
This prediction can be formulated as estimating the probability distribution over the item set $\mathcal{I}$ for the next interaction, conditioned on the historical interactions: $P(i_{n_u+1}^{(u)} = i \mid \mathcal{S}_u)$ for all $i \in \mathcal{I}$.
During training, our objective is to predict the subsequent item $i^{(u)}_{j + 1}$ for every prefix $j$ of the given sequence $\mathcal{S}_u$. For an input sequence $\mathcal{S}_u$, the desired output sequence is $[i_2^{(u)}, i_3^{(u)}, \ldots, i_{n_u +1}^{(u)}]$ \cite{kang2018self}.

\section{PRELIMINARY}
 
\subsection{Decoder-only Transformer}

The standard Transformer \cite{vaswani2017attention} comprises an encoder and a decoder. The encoder employs a bidirectional attention mechanism, allowing each token to attend to all other tokens. In contrast, the decoder uses a unidirectional attention mechanism, where each token can only attend to its preceding tokens. Consequently, decoders are often used independently for various autoregressive tasks \cite{achiam2023gpt, henighan2020scaling}.

A Transformer Decoder is constructed by stacking multiple Decoder Blocks, each primarily comprising one multi-head attention layer and one feed-forward neural network. Let the input to a Decoder Block be $\boldsymbol X \in \mathbb R^{n\times d}$, where $n$ is the sequence length and $d$ is the embedding dimension. The $i$-th row of $\boldsymbol{X}$ represents the latent vector of the $i$-th token. To ensure training stability, pre-norm is typically employed in the decoder \cite{xiong2020layer}. Denote the multi-head attention as $f_{\text{MHSA}}$ and the feedforward neural network as $f_{\text{FFN}}$. Thus, the final output of the decoder, $\boldsymbol Y$, is given by:
\begin{align}
\boldsymbol M &= \boldsymbol X + f_{\text{MHSA}}(\text{norm}(\boldsymbol X)) \\ 
\boldsymbol Y &= \boldsymbol M + f_{\text{FFN}}(\text{norm}(\boldsymbol M)) 
\end{align}
\subsubsection{Multi-Head Self-Attention} The multi-head self-attention consists of $h$ heads, each with a size of $d_h = d/h$. In each head, query, key, and value vectors are computed through projection. The attention score between any two tokens is calculated by the dot product of a token's query vector and another's key vector, followed by normalization and multiplication with the value vector to yield the output of this head, $\boldsymbol h_i$:
\begin{align}
  \boldsymbol h_i = \text{mask}\left (\varphi((\boldsymbol X\boldsymbol W_q^{(i)}) (\boldsymbol X\boldsymbol W_k^{(i)})^T) \right )(\boldsymbol X\boldsymbol W_v^{(i)})
\end{align}
where $\boldsymbol W_q^{(i)}, \boldsymbol W_k^{(i)}, \boldsymbol W_v^{(i)} \in \mathbb R^{d\times d_h}$ are learnable parameters, and $\varphi$ is the normalization function, typically softmax in NLP, while SiLU is found to be more effective in recommendation tasks \cite{zhai2024actions}. The function $\text{mask}(\boldsymbol{X})$ applies a causal mask to the matrix $\boldsymbol{X}$ by setting its upper triangular part (excluding the diagonal) to zero.
The final output of the multi-head self-attention is:
\begin{align}
f_{\text{MHSA}}(\boldsymbol X) = \text{concat}(\boldsymbol h_1, \dots, \boldsymbol h_h)\boldsymbol W_o
\end{align}
where $\boldsymbol W_o \in \mathbb R^{d\times d}$ is a learnable parameter.

\subsubsection{Feed-Forward Neural Network} In a feed-forward neural network, the input vector is first projected into a higher-dimensional space and then passed through a nonlinear activation function. This process can be represented mathematically as follows:
\begin{align}
f_{\text{FFN}}(\boldsymbol X) = \phi(\boldsymbol X \boldsymbol W_1) \boldsymbol W_2
\end{align}
where $\boldsymbol W_1 \in \mathbb R^{d\times d_{\text{FFN}}}, \boldsymbol W_2 \in \mathbb R^{d_{\text{FFN}}\times d}$ are learnable parameters, and $\phi$ denotes a non-linear activation function.

\subsection{Relative Attention Bias}

To address the inability to perceive the position of tokens when calculating attention weights, absolute or relative positional encodings are incorporated to add positional information. Relative Attention Bias (RAB), proposed in T5 \cite{raffel2020exploring}, is a method of relative positional encoding that adds a trainable bias term directly to the attention map, denoted as $\boldsymbol B = (b_{i, j})_{n\times n} \in \mathbb{R}^{n\times n}$:
\begin{align}
\boldsymbol h_i = \text{mask}(\varphi((\boldsymbol X\boldsymbol W_q^{(i)}) (\boldsymbol X\boldsymbol W_k^{(i)})^T + \boldsymbol B))(\boldsymbol X\boldsymbol W_v^{(i)})
\end{align}
The bias term is calculated using $b_{i,j} = \boldsymbol \beta_{f(i - j)}$, where $\boldsymbol \beta \in \mathbb{R}^{d_{\text{rab}}}$ represents learnable parameters, and $f$ is a bucketing function that maps the relative position to one value of $\boldsymbol \beta$. 
Similarly, we can express relative temporal information as:
\begin{align}
b_{i,j}^t = \boldsymbol \beta^t_{f_t(t_i - t_j)}
\end{align}
where, $t_i$ denotes the timestamp of the $i$-th item, and $f_t$ represents the bucketing function for relative time. In HSTU \cite{zhai2024actions}, the attention map is computed as follows:
\begin{align}
\boldsymbol A = (\boldsymbol X\boldsymbol W_q) (\boldsymbol X\boldsymbol W_k)^T + \boldsymbol B + \boldsymbol B^t
\end{align}
In \textit{FuXi}-$\alpha$ \cite{ye2025fuxi}, the relative time and position bias terms, $B$ and $B^t$, are directly applied to individual heads as the attention map. In subsequent sections, we will refer to this method as the "bucketed relative attention bias" for differentiation purposes.

\section{METHODOLOGY}

\begin{figure}
    \centering
    \includegraphics[width=0.8\linewidth]{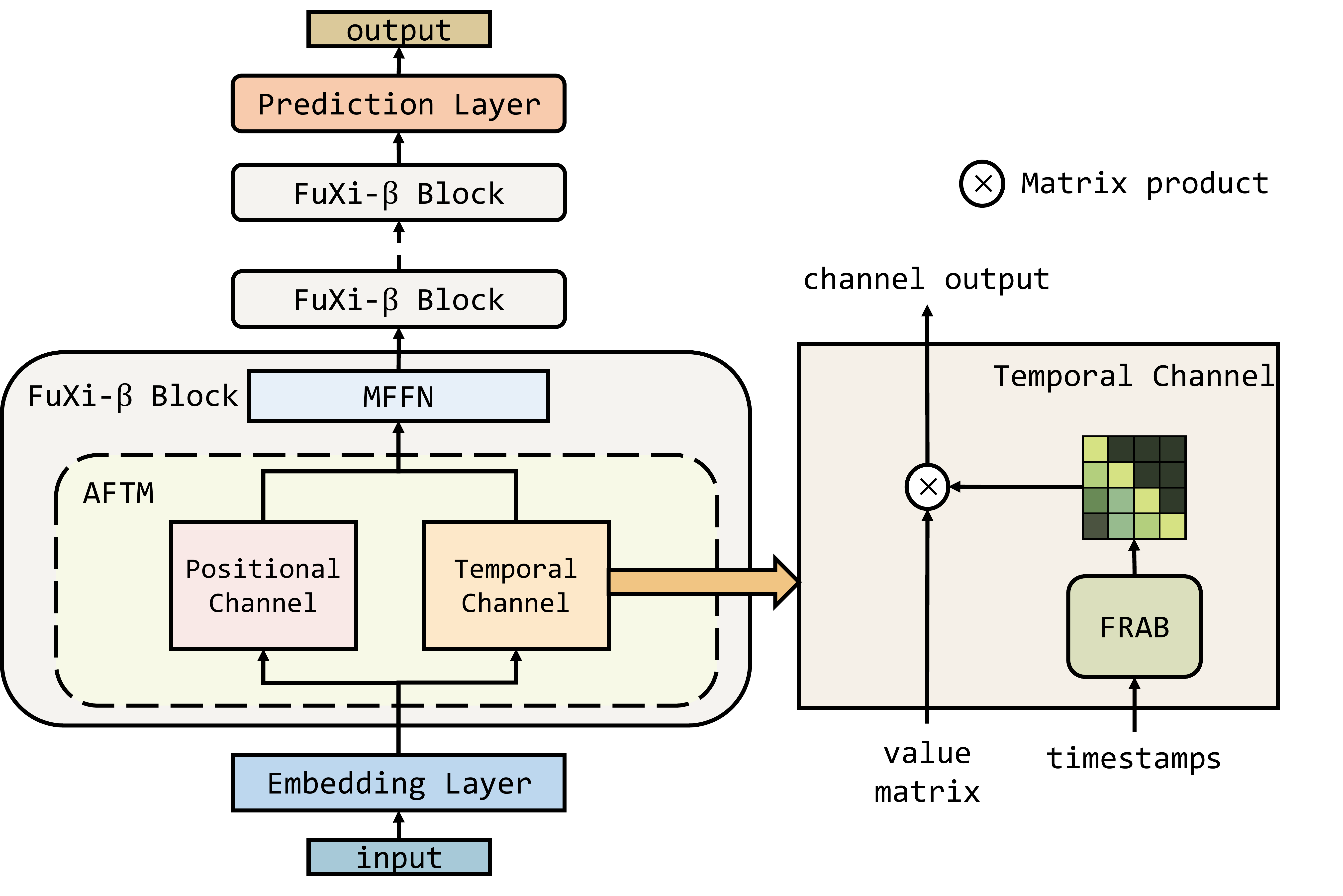}
    \caption{The overview of the proposed model FuXi-$\beta$.}
    \label{fig:structure-overview}
\end{figure}

We introduce our newly proposed framework based on the current state-of-the-art generative recommendation model, \textit{FuXi}-$\alpha$, and we refer to the new model as \textit{FuXi}-$\beta$. Although we present this framework based on \textit{FuXi}-$\alpha$, it can be applied to other Transformer-like generative recommendation models. \textit{FuXi}-$\beta$ is also a Transformer-like model, mainly consisting of $L$ layers of stacked \textit{FuXi}-$\beta$ Blocks, as shown in Figure \ref{fig:structure-overview}. In the following subsections, we will introduce \textit{FuXi}-$\beta$ in the following order: (1) an embedding layer that provides item embeddings and absolute positional information; (2) a new and fast module in the framework to model relative temporal information: the FRAB module; (3) a method within the framework to simplify the self-attention layer; (4) a discussion of other details within the \textit{FuXi}-$\beta$ Block; (5) prediction and training loss.

\subsection{Embedding Layer}

To address the variability in user interaction sequence lengths, we standardize these sequences into a fixed length $n$ through truncation or padding with a special "padding item" prior to the embedding layer. In the embedding layer, each unique item $i \in \mathcal{I}$ is mapped to a $d$-dimensional vector using a learnable embedding matrix $\mathbf{E} \in \mathbb{R}^{|\mathcal{I}| \times d}$. Additionally, we incorporate absolute positional information into each token in the embedding layer by using learnable positional encodings \cite{gehring2017convolutional}. Let the positional embedding vector for the $i$-th position be $\boldsymbol{p}_i$, and the embedding of the $i$-th item be $\boldsymbol{e}_i$. For a user $u$ with sequence $\mathcal{S}_u = \left[i_{1}^{(u)}, \ldots, i_{n_u}^{(u)}\right]$, the output of the embedding layer is expressed as 
\begin{align}
    \mathcal{S}_u = \left[\boldsymbol{e}_{1}^{(u)} + \boldsymbol{p}_1, \ldots, \boldsymbol{e}_{n_u}^{(u)} + \boldsymbol{p}_{n_u}, \boldsymbol{0}, \cdots, \boldsymbol{0}\right]
\end{align}
where $\boldsymbol{0}$ denotes the positions that are padded.

\subsection{Functional Relative Attention Bias} \label{fraw}
 The bucketed relative attention bias used for modeling relative time was inefficient due to numerous time-consuming non-continuous indexing operations. To address this inefficiency, the Functional Relative Attention Bias is proposed, which only utilizes special function and arithmetic operations to model relative temporal information. Let the timestamp of the $i$-th item be denoted as $t_i$. The attention bias for modeling the relative temporal information are computed via the following equation:
\begin{equation}
    b^t_{i,j} = f(t_j - t_i)
\end{equation}
where $f$ is a monotonically function of the following form:
\begin{equation}
    f(x) = a(1 + x)^{-b}
\end{equation}
here, $a$ and $b$ are learnable parameters. Because the given sequence is arranged in a chronological order, we assume $x$ is non-negative. To prevent numerical issues with identical timestamps, we add 1 to $x$ before applying the power function.

\subsection{Attention-Free Token Mixer} 
In the self-attention layer, beyond deriving the attention map from the query-key multiplication, the bias term from the relative attention bias module can also act as an attention map. Our ablation experiments, detailed in Section \ref{ablation_selfattention}, showed that the query-key derived attention map is unnecessary for feature interaction among items. Consequently, we removed the query and key matrices in a Transformer-like architecture and utilized the attention map from the relative attention bias module to facilitate inter-item feature interaction. We refer to this new inter-item interaction layer as the Attention-free Token Mixer (AFTM). Let $\boldsymbol X \in \mathbb{R}^{n \times d}$ denote the latent vector representation of the input items, and let $\boldsymbol t \in \mathbb{R}^{n}$ represent the sequence of timestamps corresponding to these items. Initially, the matrices $\boldsymbol U$ and $\boldsymbol V$ are computed through projection as follows:
\begin{align}
  \boldsymbol U = \phi(\boldsymbol X \boldsymbol W_u), 
  \boldsymbol V = \phi(\boldsymbol X \boldsymbol W_v)
\end{align}
where $\phi$ represents the SiLU activation function, and $\boldsymbol W_u \in \mathbb{R}^{d \times 2d}$ and $\boldsymbol W_v \in \mathbb{R}^{d \times 2d}$ are learnable parameters. Subsequently, the Relative Attention Bias (RAB) is used to transform positional information into an attention map $\boldsymbol B$, and the Functional Relative Attention Bias (FRAB) is used to convert temporal information into an attention map $\boldsymbol B^t$. The output of the AFTM is given by:
\begin{align}
    f_{\text{AFTM}}(\boldsymbol X, \boldsymbol t) = \boldsymbol U \odot \text{concat}(\boldsymbol B \boldsymbol V, \boldsymbol B^t \boldsymbol V)
\end{align}
\subsection{FuXi-$\beta$ Block}

Each \textit{FuXi}-$\beta$ Block consists of an Attention-Free Token Mixer (AFTM) and an Multistage FFN (MFFN) utilizing SwiGLU \cite{shazeer2020glu} as the activation function, and employs RMSNorm \cite{DBLP:conf/nips/ZhangS19a} as the normalization function. 
The MFFN is a variant of FFN introduced in \textit{FuXi}-$\alpha$~\cite{ye2025fuxi}.
The specific computational process is as follows:
\begin{align}
\boldsymbol{M} &= f_{\text{AFTM}}(\text{RMSNorm}(\boldsymbol{X}), \boldsymbol{t}) \\ 
\boldsymbol{f}_{\text{block}}(\boldsymbol{X}, \boldsymbol{t}) &= f_{\text{MFFN}}(\boldsymbol{M}, \boldsymbol{X}) 
\end{align}

\subsection{Prediction Layer \& Optimization objective}

Following the passage through $L$ layers of \textit{FuXi}-$\beta$ blocks, each position has garnered ample information related to the items that previously interacted. We implement a multiplication with the transpose of the input embedding matrix, which is subsequently subjected to a softmax function, thereby generating a probability distribution across the predicted items. This transformation can be mathematically articulated as follows:
\begin{align}
    P\left(i_{j}^{(u)} = i \mid (i_1^{(u)}, t_1^{(u)},1), \dots, (i_{j-1}^{(u)}, t_{j-1}^{(u)},j-1) \right) = \text{softmax}\left (\mathbf x^{B} \mathbf E^T   \right)_i
\end{align}
To expedite the training process, we incorporate the sampled softmax loss utilizing $N$ randomly selected negative samples \cite{Klenitskiy_2023}.

\subsection{Complexity Analysis}

\begin{table*}[ht]
    \centering
    \setlength{\abovecaptionskip}{-0cm}
    \setlength{\belowcaptionskip}{-0.1cm}
    \caption{Time complexity of different models} 
    \begin{tabular}{lcc}
        \toprule
        \textbf{Model} & \textbf{Computational Complexity} & \textbf{Position and Time Embedding 
        Complexity} \\
        \midrule
        \textbf{LLaMa} & $O(4nd^2 + 2n^2d + 3nd_{FFN}d)$ & $O(n)$ \text{arithmetic operations} \\
        \textbf{LLaMa-$\beta$} & $O(2nd^2 + n^2d + 3nd_{FFN}d)$ & $O(n^2)$ \text{special functions and arithmetic operations} \\
        \textbf{HSTU} & $O(5nd^2 + 2n^2d)$ & $O(n^2)$ \text{special function computation, arithmetic, and discontinuous indexing operations.} \\
        \textbf{HSTU-$\beta$} & $O(3nd^2 + n^2d)$ & $O(n^2)$ \text{special function computation and arithmetic operations} \\
        \textbf{FuXi-$\alpha$} & $O(9nd^2 + 4n^2d + 3nd_{FFN}d)$ & $O(n^2)$ \text{special function computation, arithmetic, and discontinuous indexing operations.} \\
        \textbf{FuXi-$\beta$} & $O(5nd^2 + 2n^2d + 3nd_{FFN}d)$ & $O(n^2)$ \text{special function computation and arithmetic operations} \\
        \bottomrule
    \end{tabular}
    \label{tab:complexity_analysis}
\end{table*}

In this section, we theoretically analyze the complexity variations of several Transformer-like models, namely LLaMa \cite{dubey2024llama}, HSTU \cite{zhai2024actions}, and \textit{FuXi}-$\alpha$ \cite{ye2025fuxi}, before and after applying our framework. We assume that in the self-attention layer, the condition $d_h \times h = d$ holds, where $d_h$ represents the projection dimension of each attention head, and $h$ denotes the number of attention heads. Definitions of other variables can be found in the previous sections. 
Since the discontinuous index operations involved in the bucketed relative attention bias modules are significantly more time-consuming than arithmetic operations, like multiplication and addition, we list them separately in our analysis. 
The computed complexities for different models are presented in Table \ref{tab:complexity_analysis}.

\begin{itemize}[leftmargin=*,align=left]
    \item LLaMa's self-attention requires one computation of QKV projections and output projection, with a computational cost of $4nd^2$, and one computation of attention weights, followed by multiplication with the V matrix, which incurs an additional cost of $2n^2d$. HSTU incurs an extra computational cost of $nd^2$ for calculating the U matrix. In \textit{FuXi}-$\alpha$, the dimension of the concatenated vector of all the channels is three times the embedding dimension. Therefore, this adds an extra $4nd^2$ to the complexity compared to HSTU, along with an additional $2n^2d$ due to the calculation of weights from two additional attention maps.

    \item In LLaMa-$\beta$ and HSTU-$\beta$, there is no need to compute queries and keys, nor their dot products, thereby reducing the complexity by $2nd^2 + n^2d$ compared to their respective base models. In \textit{FuXi}-$\beta$, the dimension of the concatenated vector is reduced, resulting in a computational reduction of $4nd^2 + 2n^2d$ compared to \textit{FuXi}-$\alpha$.

    \item In both HSTU and \textit{FuXi}-$\alpha$, the application of FRAB can reduce time consumption by avoiding discontinuous index operations. However, LLaMa uses RoPE \cite{su2024roformer} to utilize positional information, and replacing it with FRAB would increase computational complexity. Nonetheless, this substitution addresses the issue that RoPE encoding cannot simultaneously utilize both temporal and positional information.
\end{itemize}

In a word, our method can reduce approximately half of the computational load in the self-attention layer and replace the high time-consuming indexing operations in the bucketed relative attention bias with more efficient calculations.

\section{EXPERIMENTS}\label{experiment}

\begin{table}[t]
\setlength{\abovecaptionskip}{-0cm}
\setlength{\belowcaptionskip}{-0.1cm}
 \caption{Dataset statistics.}
 \centering
 	\setlength{\tabcolsep}{1mm}
 \begin{tabular}{@{} c|c|c|c|c @{}}
 \hline
 \textbf{Dataset} 	  & \textbf{User}   & \textbf{Item} & \textbf{Interactions} & \textbf{Avg. Len.}   \\
 \hline
 MovieLens-1M  & 6,041 & 3,706 & 1,000,209 & 165.60   \\
 MovieLens-20M & 138,493 & 26,744 & 20,000,263 & 144.41   \\
 Daily Recomm. & 19,252,028 & 234,488 & 1,023,711,774 & 53.17 \\ 
 All scenarios & 28,927,689 & 476,544 & 1,313,729,225 & 40.89 \\ 
 \hline
\end{tabular}
\label{tab:dataset_statistics}
\end{table}

\begin{table*}[t]
\setlength{\abovecaptionskip}{0cm}
\setlength{\belowcaptionskip}{0cm}
\centering
\caption{The overall performance comparison on public datasets. 
We use bold to indicate the best result and underline to indicate the second best result. We set the training time of SASRec as 1.0. The training times of other models are expressed as multiples of SASRec's training time.}
\setlength{\tabcolsep}{1mm}{
\small
\begin{tabular}{c|c|c|c|c|c|c|c|c|c|c|c|c}
\midrule[0.25ex]
\textbf{Dataset} & \multicolumn{6}{c|}{\textbf{MovieLens-1M}} & \multicolumn{6}{c}{\textbf{MovieLens-20M}} \\ \hline 
\textbf{Model} & NDCG@10 & NDCG@50 & HR@10 & HR@50 & MRR & Time &  NDCG@10 & NDCG@50 & HR@10 & HR@50 & MRR & Time \\ \hline \hline
\textbf{BPRMF} & 0.0607 & 0.1027 & 0.1185 & 0.3127 &  0.0556 & - & 0.0629 & 0.1074 & 0.1241 & 0.3300 & 0.0572 & -  \\
\textbf{GRU4Rec} & 0.1015 & 0.1460 & 0.1816 & 0.3864 & 0.0895 & - & 0.0768 & 0.1155 & 0.1394 & 0.3177 & 0.0689  & -  \\
\textbf{NARM} & 0.1350 & 0.1894 & 0.2445 & 0.4915 & 0.1165 & - & 0.1037 & 0.1552 & 0.1926 & 0.4281 & 0.0910  & - \\

\hline 
 \textbf{FLASH}& 0.1573 &0.2144 & 0.2830 &0.5426 &0.1341 & 1.022 &0.1496&0.2057&0.2682 & 0.5226 & 0.1285 & 1.210\\
 \textbf{RetNet}&0.1601&0.2184& 0.2925 &0.5557&0.1352& 0.975 &0.1687&0.2249&0.2979 & 0.5526 & 0.1442 & 1.034 \\
\hline
\textbf{SASRec} & 0.1594 & 0.2187 &0.2824 & 0.5500 & 0.1375 & 1.000 & 0.1553 & 0.2119 & 0.2781 & 0.5353 & 0.1330 & 1.000 \\
\textbf{LLaMa} & 0.1620 & 0.2207 & 0.2926 & 0.5591 & 0.1373 & 0.978 & 0.1640 & 0.2206& 0.2915&0.5476 & 0.1402 & 1.006 \\
\textbf{HSTU} & 0.1639 & 0.2238 & 0.2969 & 0.5672 & 0.1390 & 1.019 & 0.1642 & 0.2225 & 0.2909 & 0.5553 & 0.1410 & 1.072 \\
\textbf{FuXi-$\alpha$} & 0.1835 & 0.2429 & 0.3254 & 0.5941 &0.1557 & 1.078 & 0.1954 & 0.2533 & 0.3353 & 0.5969 & 0.1677 & 1.126 \\
\textbf{FuXi-$\beta$} &0.1848 & 0.2432 &0.3231 & 0.5866 & 0.1578 & 1.019 &0.1944 & 0.2513 &0.3325 &0.5899 & 0.1671 & 1.006 \\
\hline

\textbf{SASRec-Large} & 0.1186 & 0.1733 & 0.2183 & 0.4671 & 0.0186 & 1.000 & 0.0206 & 0.0379 & 0.0412 & 0.1209 & 0.0207 & 1.000 \\

\textbf{LLaMa-Large} & 0.1659 & 0.2257 & 0.2990 & 0.5692 & 0.1408 &  0.989 & 0.1842 & 0.2412 & 0.3202 & 0.5776 & 0.1576 & 1.022 \\
\textbf{HSTU-Large} & 0.1844 & 0.2437 & 0.3255 & 0.5929 & 0.1568  & 1.134 & 0.1995 & 0.2572 & 0.3407 & 0.6012 & 0.1714 & 1.187  \\
\textbf{FuXi-$\alpha$-Large}  & \underline{0.1934} &  \underline{0.2518} & \underline{0.3359} & \underline{0.5983} & \textbf{0.1651} & 1.230 & \underline{0.2086} & \underline{0.2658} & \underline{0.3530} & \textbf{0.6113} & \underline{0.1792} & 1.371  \\

\textbf{FuXi-$\beta$-Large} & \textbf{0.1947} & \textbf{0.2523} & \textbf{0.3428} & \textbf{0.6022} & \underline{0.1645} & 1.068 & \textbf{0.2117} & \textbf{0.2677} & \textbf{0.3566} & \underline{0.6095} & \textbf{0.1818} & 1.000 \\
\hline
\hline
\end{tabular}
}
\label{tab:public_performance}
\end{table*}

\begin{table*}[t]
    \setlength{\abovecaptionskip}{-0.0cm}
    \setlength{\belowcaptionskip}{-0.2cm}
    \centering
    \caption{Performance comparison on Industrial dataset. }
    \setlength{\tabcolsep}{1mm}{
    \small
    \begin{tabular}{c|c|c|c|c|c|c|c|c|c|c|c|c}
    \midrule[0.25ex]
    \textbf{Dataset} &
    \multicolumn{6}{c|}{\textbf{Daily Recommendations}} & \multicolumn{6}{c}{\textbf{All scenarios}} \\ \hline 
    \textbf{Model} & NDCG@10 & NDCG@50 & HR@10 & HR@50 & MRR & Time & NDCG@10 & NDCG@50 & HR@10 & HR@50 & MRR & Time \\\hline \hline
    
    \textbf{SASRec} & 0.0795 & 0.1488 & 0.1696 & 0.4887 & 0.0707 & 1.000 & 0.1638 & 0.2263 & 0.2971 & 0.5767 & 0.1399 & 1.000
  \\
    \textbf{LLaMa} & 0.1486 & 0.2205 &  0.2881 & 0.6139 & 0.1248 & 1.208 & 0.2229 & 0.2795 & 0.3732 & 0.6259 & 0.1919 & 1.212
 \\
    \textbf{HSTU} & 0.1516 & 0.2235 & 0.2921 & 0.6183 & 0.1274 & 1.153 & 0.2295 & 0.2859 & 0.3838 & 0.6355 &0.1972  & 1.144  \\
    \textbf{FuXi-$\alpha$}  & \underline{0.1785} & \underline{0.2488} & \underline{0.3271} & \underline{0.6449} & \underline{0.1513} & 1.356
 & \underline{0.2493} & \underline{0.3039} & \underline{0.4080} & \underline{0.6512} & \underline{0.2150} & 1.323
 \\ 
     \textbf{FuXi-$\beta$}  & \textbf{0.2631}	&  \textbf{0.3134}	&  \textbf{0.4228}	&  \textbf{0.6497} &  \textbf{0.2269}  & 1.142 & \textbf{0.3172}
& \textbf{0.3631} & \textbf{0.4740} & \textbf{0.6781} & \textbf{0.2806} & 1.050
     \\ \hline
    \end{tabular}}
    \label{tab:industrial_performance}
\end{table*}

\begin{table*}[t]
    \setlength{\abovecaptionskip}{-0.0cm}
    \setlength{\belowcaptionskip}{-0.2cm}
    \centering
    \caption{Ablation study results on the public datasets}
    \setlength{\tabcolsep}{1mm}{
    \small
    \begin{tabular}{c|c|c|c|c|c|c|c|c|c|c|c|c}
    \midrule[0.25ex]
    \textbf{Dataset} &
    \multicolumn{6}{c|}{\textbf{MovieLens-1M}} & \multicolumn{6}{c}{\textbf{MovieLens-20M}} \\ \hline 
    \textbf{Model} & NDCG@10 & NDCG@50 & HR@10 & HR@50 & MRR & Time & NDCG@10 & NDCG@50 & HR@10 & HR@50 & MRR & Time \\\hline \hline
    \textbf{FuXi-$\alpha$} & 0.1835 & 0.2429 & 0.3254 & 0.5941 &0.1557 & 1.000 & 0.1954 & 0.2533 & 0.3353 & 0.5969 & 0.1677 & 1.000 \\
    \textbf{FuXi-$\beta$-FRAB} & 0.1823 & 0.2407& 0.3244 & 0.5876 & 0.1541 & 0.997 & 0.1963 & 0.2530 & 0.3355 & 0.5923 & 0.1685 & 0.962 \\
    \textbf{FuXi-$\beta$-AFTM} & 0.1797 & 0.2382 &0.3188 &0.5819 &0.1526& 0.994 & 0.1956 & 0.2534 & 0.3352 & 0.5964 & 0.1679 & 0.959 \\
    \textbf{FuXi-$\beta$} &0.1848 & 0.2432 &0.3231 & 0.5866 & 0.1578 & 0.945 &0.1944 & 0.2513 &0.3325 &0.5899 & 0.1671 & 0.894 \\
    \hline 
    \textbf{FuXi-$\alpha$-Large}  & 0.1934 &  0.2518 & 0.3359 & 0.5983 & 0.1651 & 1.000 & 0.2086 & 0.2658 & 0.3530 & 0.6113 & 0.1792 & 1.000  \\ 
    \textbf{FuXi-$\beta$-FRAB-Large} & 0.1871 & 0.2454 & 0.3327 & 0.5944 & 0.1578 & 0.913 & 0.2097 & 0.2658 & 0.3532 & 0.6062 & 0.1803 & 0.867 \\
    \textbf{FuXi-$\beta$-AFTM-Large} & 0.1923 & 0.2507 & 0.3407 & 0.6030 & 0.1624 & 0.982 & 0.2107 & 0.2672 & 0.3550 & 0.6098 &  0.1811 & 0.858 \\
    \textbf{FuXi-$\beta$-Large} & 0.1947 & 0.2523 & 0.3428 & 0.6022 & 0.1645 & 0.869 & 0.2117 & 0.2677 & 0.3566 & 0.6095 & 0.1818 & 0.730 \\
      \hline
    \end{tabular}}
    \label{tab:ablation_performance}
\end{table*}

\begin{table}[t]
    \setlength{\abovecaptionskip}{-0.0cm}
    \setlength{\belowcaptionskip}{-0.2cm}
    \centering
    \caption{Comparison of the performance of FRAB using different functions.}
    \setlength{\tabcolsep}{1mm}{
    \small
    \begin{tabular}{c|c|c|c|c|c|c}
    \midrule[0.25ex]
    \textbf{Dataset} &
    \multicolumn{3}{c|}{\textbf{MovieLens-1M}} & \multicolumn{3}{c}{\textbf{MovieLens-20M}} \\ \hline
    \textbf{Function} & NG@50 & HR@50 & MRR & NG@50 & HR@50 & MRR \\ \hline \hline
    \textbf{linear}  & 0.2120 & 0.5345 & 0.1330 & 0.2488 & 0.5869 & 0.1646 \\
    \textbf{log}     & 0.2350 & 0.5817 & 0.1487 & 0.2617 & 0.6032 & 0.1761 \\
    \textbf{exp}     & 0.2500 & 0.5959 & 0.1635 & 0.2658 & 0.6057 & 0.1804 \\
    \textbf{sin}     & 0.2367 & 0.5771 & 0.1521 & 0.2526 & 0.5870 & 0.1690 \\
    \textbf{mixed}   & 0.2136 & 0.5378 & 0.1336 & 0.2498 & 0.5844 & 0.1665 \\
    \textbf{NN}      & 0.2411 & 0.5806 & 0.1565 & 0.2514 & 0.5819 & 0.1690 \\
    \textbf{zero}    & 0.2345 & 0.5776 & 0.1490 & 0.2434 & 0.5765 & 0.1605 \\ 
    \textbf{bucket}  & 0.2507 & 0.6030 & 0.1624 & 0.2672 & 0.6098 & 0.1811 \\
    \textbf{pow}     & 0.2523 & 0.6022 & 0.1645 & 0.2677 & 0.6095 & 0.1818 \\ 
    \hline
    \end{tabular}}
    \label{tab:function_performance}
\end{table}

\begin{table*}[t]
    \setlength{\abovecaptionskip}{-0.0cm}
    \setlength{\belowcaptionskip}{-0.2cm}
    \centering
    \caption{Results of the ablation study on the self-attention layer}
    \setlength{\tabcolsep}{1mm}{
    \small
    \begin{tabular}{c|c|c|c|c|c|c|c|c|c|c|c|c}
    \midrule[0.25ex]
    \textbf{Dataset}  &
    \multicolumn{6}{c|}{\textbf{MovieLens-1M}} & \multicolumn{6}{c}{\textbf{MovieLens-20M}} \\ \hline 
    \textbf{Model} & NDCG@10 & NDCG@50 & HR@10 & HR@50 & MRR & Time & NDCG@10 & NDCG@50 & HR@10 & HR@50 & MRR & Time \\\hline \hline
    \textbf{FuXi-$\beta$-FRAB-Large} & 0.1871 & 0.2454 & 0.3327 & 0.5944 & 0.1578 & 1.000 & 0.2097 & 0.2658 & 0.3532 & 0.6062 & 0.1803 & 1.000\\
   \textbf{- query-key attention map} & \textbf{0.1947} & \textbf{0.2523} & \textbf{0.3428} & \textbf{0.6022} & \textbf{0.1645} & \textbf{0.951} & \textbf{0.2117} & \textbf{0.2677} & \textbf{0.3566} & \textbf{0.6095} & \textbf{0.1818} & \textbf{0.842} \\
     \textbf{- positional attention map} & 0.1800 & 0.2393 & 0.3164 & 0.5833 & 0.1539 & 0.951 & 0.2093 & 0.2659&0.3527&0.6078&0.1801& 0.946\\
    \textbf{- temporal attention map} &0.1741&0.2329&0.3096&0.5750&0.1480& 0.990 &0.1863 & 0.2418 & 0.3218 & 0.5730 & 0.1594 & 0.914 \\
    
      \hline
    \end{tabular}}
    \label{tab:ablation_attention}
\end{table*}

\begin{table}[t]
    \setlength{\abovecaptionskip}{-0.0cm}
    \setlength{\belowcaptionskip}{-0.2cm}
    \centering
    \caption{Compatibility analysis on the public datasets. We express the training time as a multiple of the training time for SASRec with the same parameter settings.}
    \setlength{\tabcolsep}{1mm}{
    \small
    \begin{tabular}{c|c|c|c|c|c|c}
    \midrule[0.25ex]
    \textbf{Dataset} &
    \multicolumn{3}{c|}{\textbf{MovieLens-1M}} & \multicolumn{3}{c}{\textbf{MovieLens-20M}} \\ \hline 
    \textbf{Model} & \text{NG@50} & \text{HR@50} & \text{Time} & \text{NG@50} & \text{HR@50} & \text{Time} \\ \hline \hline \textbf{LLaMa-Large} & 0.2257 & 0.5692 & 0.989 & 0.2412 & 0.5776 & 1.022 \\ \textbf{LLaMa-$\beta$-Large} & 0.2354 & 0.5785 & 1.016 & 0.2560 & 0.6006 & 0.902 \\ \hline \textbf{HSTU-Large} & 0.2437 & 0.5929 & 1.134 & 0.2572 & 0.6012 & 1.187 \\ \textbf{HSTU-$\beta$-Large} & 0.2425 & 0.5828 & 0.959 & 0.2520 & 0.5940 & 0.851 \\ \hline \textbf{FuXi-$\alpha$-Large} & 0.2518 & 0.5983 & 1.230 & 0.2658 & 0.6113 & 1.371 \\ \textbf{FuXi-$\beta$-Large} & 0.2523 & 0.6022 & 1.068 & 0.2677 & 0.6095 & 1.000 \\ \hline
    

      \hline
    \end{tabular}}
    \label{tab:compatibility_analysis}
\end{table}

\subsection{Experiment Setup}\label{ExperimentSetup}

\subsubsection{Datasets}

We conducted experiments on two public datasets (MovieLens-1M and MovieLens-20M) and two private large-scale industrial datasets (Daily Recommendations and All Scenarios). The details of these datasets are outlined below:

\begin{itemize}[leftmargin=*,align=left]
    \item \textbf{MovieLens}\footnote{https://grouplens.org/datasets/movielens/}. MovieLens is a popular benchmark dataset for movie recommendation. The dataset comprises various subsets of different sizes. In our work, we use the MovieLens-1M and MovieLens-20M subsets for our experiment.
    \item \textbf{Daily Recommendations}. In the daily recommendation scenario, 30 exclusive songs are selected for each user every day.
    The displayed list is updated in a daily manner.
    We sampled a portion of the data to form an offline dataset, which includes tens of millions of users and billions of interactions.
    \item \textbf{All Scenarios}. The all scenarios dataset includes user behaviors from all music recommendation scenarios, with a more complex and challenging data distribution, which can better reflect the effectiveness and robustness of the model. 
    We also sampled tens of millions of users and billions of interactions to build an offline dataset.
\end{itemize}

The statistical information about the datasets is presented in Table \ref{tab:dataset_statistics}. For the public datasets, we employed the same preprocessing techniques and the train/validation/test set partitioning method as used in HSTU \cite{zhai2024actions} and \textit{FuXi}-$\alpha$ \cite{ye2025fuxi}. 
For the industrial datasets, we applied similar preprocessing methods.

\subsubsection{Baselines}
We conducted a validation of \textit{FuXi}-$\beta$ in two aspects: performance and efficiency. For performance, we employed traditional models (BPRMF \cite{rendle2012bpr}, GRU4Rec \cite{hidasi2015session}, and NARM \cite{li2017neural}) as well as autoregressive generative models (SASRec \cite{kang2018self}, LLaMa \cite{dubey2024llama}, HSTU \cite{zhai2024actions}, and \textit{FuXi}-$\alpha$ \cite{ye2025fuxi}) as baselines. Regarding efficiency, we also introduced some work on efficient transformers in NLP, including FLASH \cite{hua2022transformer}, and RetNet \cite{sun2023retentive}.

\subsubsection{Evaluation Metrics}
We employ three widely used metrics (NDCG@$K$, HR@$K$, and MRR) to evaluate the recall quality of the model. Higher values of these metrics indicate better performance. We rank all items based on the model's results and report the outcomes for $K = 10$ and $K = 50$ by default. Additionally, we measure the efficiency of the model using the total training time; a shorter training time implies a faster model.

\subsubsection{Parameter Settings}
We implemented our model, \textit{FuXi}-$\beta$, using PyTorch \cite{paszke2019pytorch}, and employed the Accelerate library \cite{kotter2012accelerate} to achieve multi-NPU and multi-machine parallel training. 
In public datasets, we adopted the same parameters as HSTU \cite{zhai2024actions}, except for the number of layers. For models with a Feedforward Neural Network (FFN), the hidden layer width is the same as the embedding dimension in MovieLens-1M, while it is four times the embedding dimension in MovieLens-20M. In experiments on public datasets, we set the number of layers for the base model to 2, whereas the Large model refers to the corresponding model with 8 layers. For industrial datasets, we set the embedding size to 256 and the number of layers to 4 for all models to maintain consistency with the online version.

\subsection{Performance Comparison}\label{PerformanceComparison}
\subsubsection{Public Dataset Performance}

The results of \textit{FuXi}-$\beta$ and the baseline models on the public datasets are presented in Table \ref{tab:dataset_statistics}. 
We observe the following:
\begin{itemize}[leftmargin=*,align=left]
    \item Generative recommendation models generally outperform traditional models, even when the model has only 2 layers.
    \item As an early model, SASRec's structural design may lead to training difficulties. Under our training parameters, increasing the layers to 8 actually resulted in a decline in performance. However, other generative models have successfully increased their layers to 8. 
    \item Although FLASH outperforms the Transformer in NLP, it performs poorly in recommendation tasks. RetNet, despite having equally performance compared to models like LLaMa and HSTU, still lags behind the strongest baseline, \textit{FuXi}-$\alpha$, due to its lack of adaptation to recommendation tasks.
    \item \textit{FuXi}-$\beta$ generally exhibits performance similar to \textit{FuXi}-$\alpha$. When using a 2-layer model, its performance on MovieLens-20M slightly decreases, while in other cases, it shows a modest improvement.
\end{itemize}

\subsubsection{Industrial Dataset Performance}

Table \ref{tab:industrial_performance} shows 
the performance comparison of our proposed \textit{FuXi}-$\beta$ against several baseline models using two large-scale private industrial datasets. Our \textit{FuXi}-$\beta$ model outperforms both \textit{FuXi}-$\alpha$ and HSTU across two datasets. In the first dataset, the NDCG@10 metric has improved by 47.4\% and 73.5\% respectively, compared to \textit{FuXi}-$\alpha$ and HSTU. In the second dataset, the improvements are 27.2\% and 38.2\% respectively. This may be attributed to the fact that industrial datasets are more complex than public datasets. The incorporation of queries and keys in the self-attention layer, along with bucketed relative attention bias, increases the model's complexity, making it prone to overfitting. However, Our framework mitigates the issue of overfitting, resulting in an exceptional performance on the industrial datasets.

\begin{figure}
    \centering
    \setlength{\abovecaptionskip}{0pt}
    \setlength{\belowcaptionskip}{-15pt}
        \includegraphics[width=0.8\linewidth]{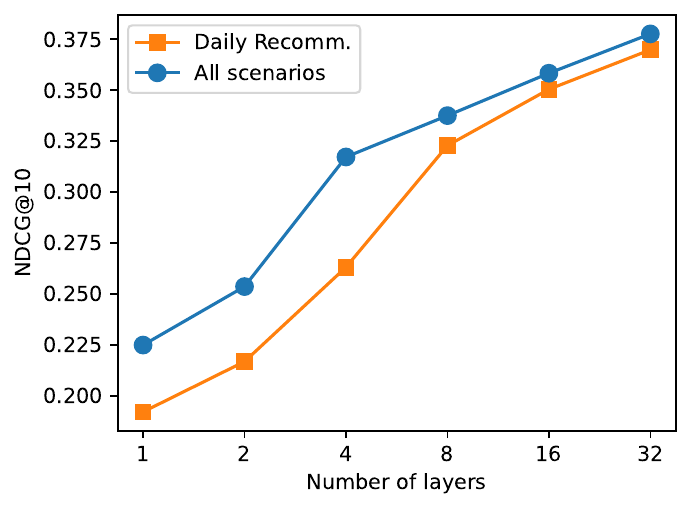}
    \caption{Scaling of FuXi-$\beta$ on Industrial Dataset.}
    \label{fig:industrial_scaling}
\end{figure}

\subsubsection{Scaling on Industrial Dataset}

We investigated the performance changes of \textit{FuXi}-$\beta$ as the model size increases on both industrial datasets, with the results shown in Figure~\ref{fig:industrial_scaling}. 
Due to memory constraints, we only increase the number of layers up to 32 layers. 
Additionally, the performance of \textit{FuXi}-$\beta$ continuously improves as the number of layers increases. This illustrates that \textit{FuXi}-$\beta$ exhibits excellent performance on large-scale datasets while still adhering to the scaling law, which states that the model's performance continues to improve with the increase in data, model size, and computational resource.

\subsection{Efficiency Comparison}\label{EfficiencyComparison}

\subsubsection{Public Datasets.} Table \ref{tab:public_performance} presents the total training time for different models over the same number of epochs. Due to the significant structural and performance differences between traditional models and generative recommendation models, we do not consider the efficiency of these traditional models. 
Although FLASH have a lower complexity~\cite{hua2022transformer}, it do not exhibit a noticeable speed advantage when the sequence length is short. RetNet exhibits relatively fast training speeds. However, due to its lack of adaptation for recommendation problems, its performance still falls short compared to some recommendation models.
The training times for LLaMa, HSTU, and \textit{FuXi}-$\alpha$ are gradually increasing as the models become more complex. \textit{FuXi}-$\beta$, when using 2 layers, reduces the training time by 5.51\% and 10.6\% on two datasets, respectively, compared to \textit{FuXi}-$\alpha$. When using 8 layers, the reductions in training time are 13.1\% and 27.0\%, respectively.

\subsubsection{Industrial Datasets.} Table \ref{tab:industrial_performance} presents the training times of different models on industrial datasets. For LLaMa, we employed the same relative attention bias as HSTU instead of RoPE~\cite{su2024roformer}, which results in a longer training time on industrial datasets compared to HSTU.
\textit{FuXi}-$\alpha$, due to its more complex architecture, lagged behind LLaMa in terms of speed. In contrast, \textit{FuXi}-$\beta$, by applying our new framework to simplify its structure, reduced training time by 15.8\% and 20.8\% on the two datasets compared to \textit{FuXi}-$\alpha$, respectively.
 
\subsection{Ablation Study of FuXi-$\beta$}

\subsubsection{Component Ablation Study} We first investigated the impact of different components on the model. We denote \textit{FuXi}-$\beta$-FRAB as the model that replaces the RAB module in \textit{FuXi}-$\alpha$ with the FRAB module to utilize relative temporal information, and the model that removes all query and key matrices from \textit{FuXi}-$\alpha$ as \textit{FuXi}-$\beta$-AFTM. The experimental results on public datasets are shown in Table \ref{tab:ablation_performance}. Although using FRAB alone and removing queries and keys led to a certain degree of performance degradation on MovieLens-1M, the combination of these two modules still achieved performance close to that of \textit{FuXi}-$\alpha$. On MovieLens-20M, these modifications resulted in a slight performance enhancement. On MovieLens-1M, they reduced the training time for the Large model by 8.68\% and 1.78\%, respectively, while on MovieLens-20M, they reduced the training time by 13.3\% and 14.2\%, respectively. This indicates that in smaller models, the acceleration is mainly due to FRAB, whereas in larger models, both contribute significantly to the acceleration.

\begin{figure*}[t]
    \centering    
    \subfigure[Plots on MovieLens-1M]{
        \label{fig:attention_weights_a}     
        \includegraphics[width=0.30\linewidth]{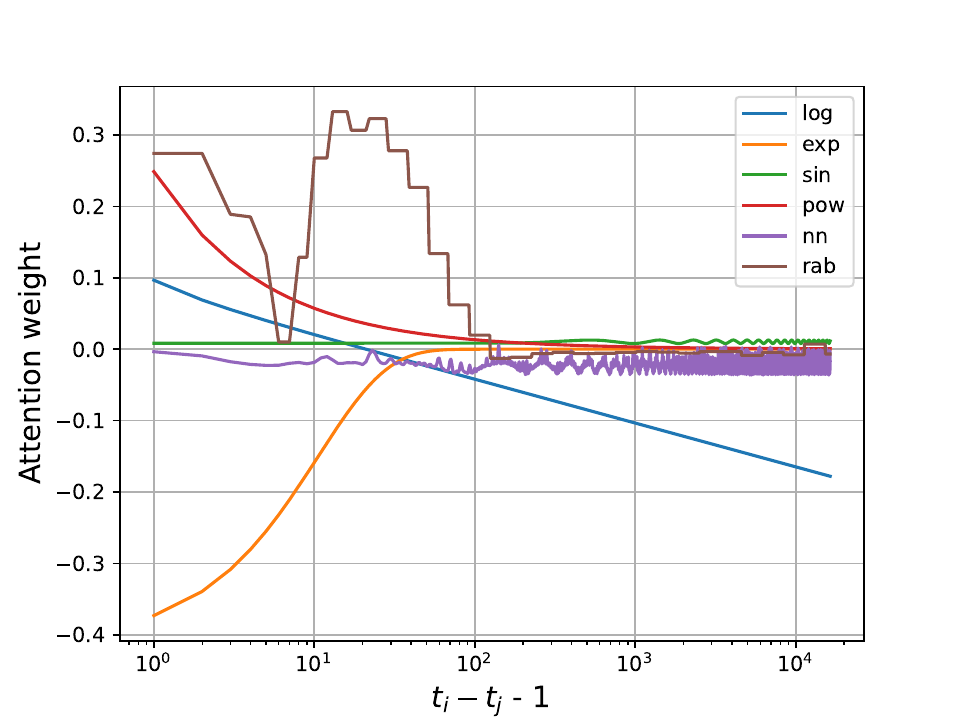}  
    }%
    \subfigure[Plots on MovieLens-20M]{
        \label{fig:attention_weights_b}     
        \includegraphics[width=0.30\linewidth]{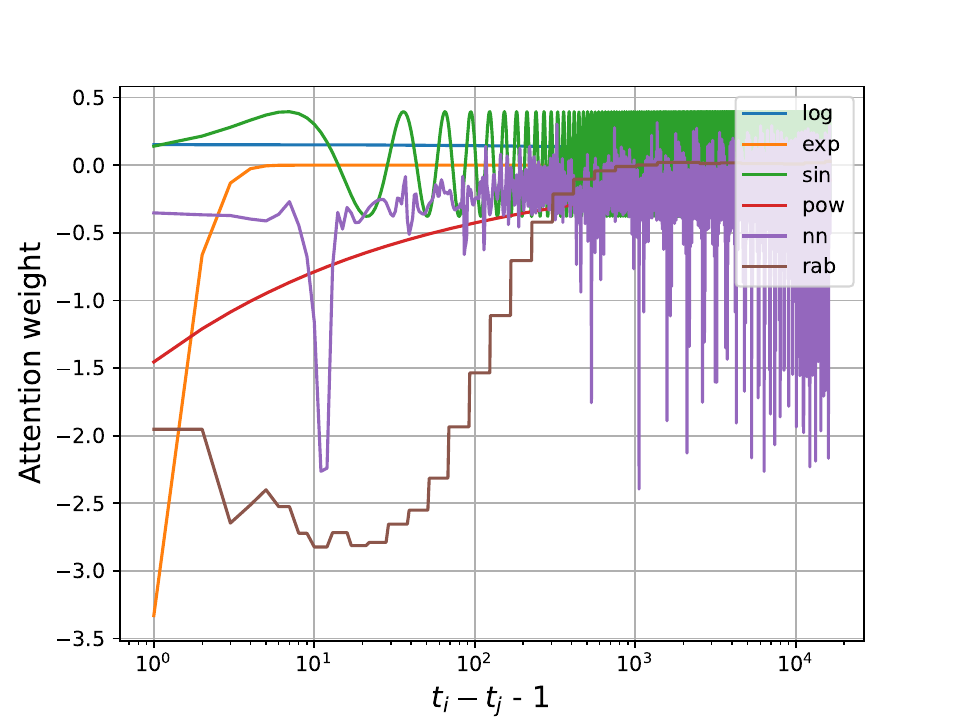}  
    }%
    \subfigure[Plots on MovieLens-20M] { 
        \label{fig:attention_weights_c}     
        \includegraphics[width=0.30\linewidth]{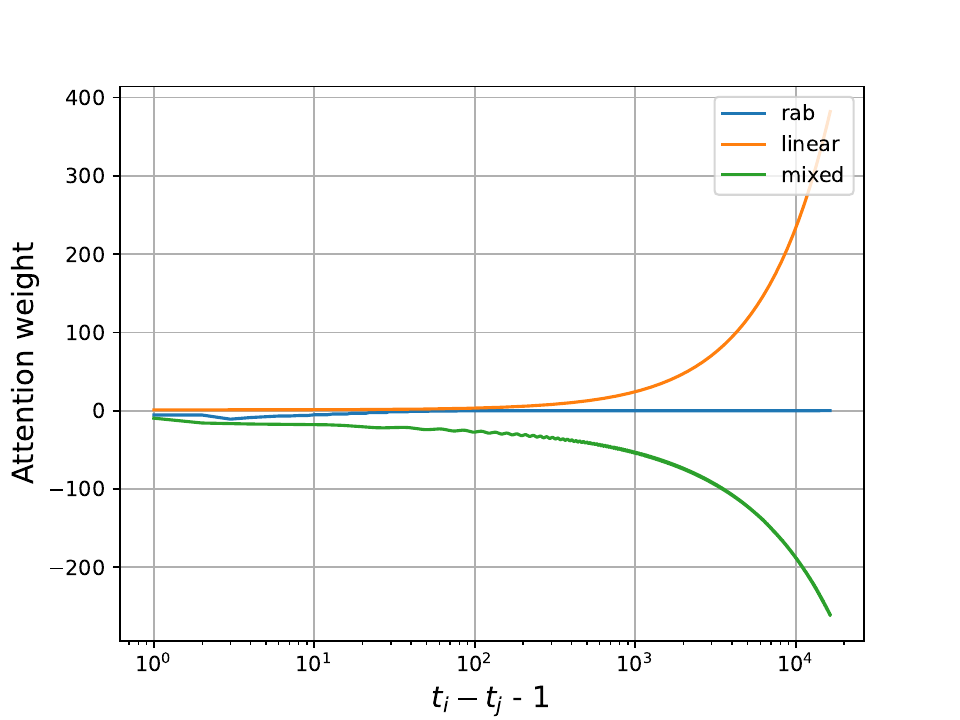}     
    }    
    \caption{Variation of attention weights produced by the FRAB module with respect to relative time using different functions.}    
    \label{fig:attention_weights}   
\end{figure*}

\subsubsection{Function Variations on Functional Relative Attention Bias} To validate the effectiveness of the function employed in our FRAB module, we conducted experiments by replacing it with various alternative functions. The functions considered for this purpose are as follows:

\begin{enumerate}[leftmargin=*,align=left]
    \item Linear function: $f_{\text{lin}}(x) = a \cdot x + b$
    \item Logarithmic function: $f_{\text{log}}(x) = a \cdot \log(1 + \exp(b) \cdot x) + c$
    \item Exponential function: $f_{\text{exp}}(x) = a \cdot \exp(-\exp(b)\cdot  x)$
    \item Sine function: $f_{\text{sin}}(x) = c \cdot \sin(a \cdot x + b) + d$
    \item Power function: $f_{\text{pow}}(x) = a \cdot (1 + x)^{-b}$, which is used in our module
    \item Mixed function: $f_{\text{mix}}(x) = \frac{f_\text{lin}(x) + f_\text{log}(x) + f_\text{exp}(x) + f_\text{sin}(x) + f_\text{pow}(x)}{5}$
    \item A small 3-layer Multi-Layer Perceptron (MLP) $f_{\text{nn}}$, utilizing sinusoidal functions and SiLU as activation functions.
    \item Zero function: $f_{\text{zero}}(x) = 0$
    \item Bucket function: $f_{\text{bucket}}$ represents the method used in relative attention bias \cite{zhai2024actions}.
\end{enumerate}

In the above functions, $a$, $b$, $c$, and $d$ represent learnable parameters unique to each function. Our experimental results on MovieLens-1M are presented in Table \ref{tab:function_performance}. From the results, replacing the power function we used with other functions results in varying degrees of performance degradation.

To further investigate the effects of different functions, we visualize the attention weights in the temporal channel with the time difference as the horizontal axis and the attention weight as the vertical axis. The results of the plot are shown in Figure \ref{fig:attention_weights}. We have the following observations: 

\begin{itemize}[leftmargin=*,align=left]
    \item The function $f_{\sin}$ varies periodically at a fixed frequency, but it is not effective in distinguishing between items that are far apart and those that are close together.
    \item The functions $f_{\text{mixed}}$ and $f_{\text{NN}}$ are difficult to train due to their complex models, resulting in suboptimal performance.
    \item The functions $f_{\text{lin}}$ and $f_{\text{log}}$ fail to adequately control the numerical range, leading to unreasonable weights assigned to items that are far apart, thus resulting in poor performance.
    \item The functions $f_{\text{pow}}$ and $f_{\text{exp}}$ have absolute values that monotonically decrease, which assigns higher attention weights to recent items. Additionally, the rate of decrease slows down, making it easier to distinguish recent features, leading to good performance. Since $f_{\text{pow}}$ can be expressed as $ae^{-b\ln(1+x)}$, it has a slower decay rate compared to $f_{\text{exp}}$. Therefore, $f_{\text{pow}}$ is more effective at assigning weights to items over longer distances, which is why it performs slightly better than $f_{\text{exp}}$.
\end{itemize}

\vspace{3pt}
\subsubsection{Ablation Study of Simplifying the Self-Attention Layer}
\label{ablation_selfattention}

To investigate whether there is redundancy among the three types of attention maps (query-key, temporal, and positional) in the self-attention layer for recommendation tasks, we empirically conducted ablation experiments based on \textit{FuXi}-$\beta$-FRAB-Large. The results are shown in Table \ref{tab:ablation_attention}. Removing the temporal attention map significantly degrades the model's performance. Removing the positional attention map leads to a performance drop on MovieLens-1M. In contrast, removing the query-key attention map not only results in the fastest execution time but also achieves the best performance on both datasets.

\vspace{5pt}
\subsubsection{Compatibility Analysis} 
\label{compatibility_analysis}
We validated our proposed framework across multiple models. Since RoPE used by LLaMa requires multiplying the query and key vectors to encode relative positional information, which is incompatible with the AFTM module, we replaced it with relative attention bias in LLaMa-$\beta$. In addition, we applied the FRAB module to process timestamps and removed the query and key matrices. In HSTU-$\beta$, we made modifications in a manner similar to \textit{FuXi}-$\beta$. The results are shown in Table \ref{tab:compatibility_analysis}. For LLaMa, due to the application of fusion operators for acceleration and the low time complexity of RoPE, the training time for LLaMa-$\beta$ on the MovieLens-1M dataset increased. However, on the MovieLens-20M dataset, the size of the query and key matrices increased so their removal can significantly reduce the training time, resulting in an 11.8\% decrease for LLaMa-$\beta$. The performance of LLaMa-$\beta$ was significantly improved due to the addition of temporal information. For HSTU-$\beta$, the training time was reduced by 15.5\% and 28.3\% on the two datasets, respectively, but its performance significantly declined. This decline may be attributed to HSTU's weakened implicit feature interaction, which hinders the effective utilization of information obtained from explicit feature interactions. For \textit{FuXi}-$\alpha$, the training time was reduced by 13.1\% and 27.0\% on the two datasets, respectively, with a slight improvement in accuracy.

\section{CONCLUSION}

In our work, we focus on simplifying generative recommendation models, and we propose a new framework. On the one hand, we introduce the Functional Relative Attention Bias module, which employs special functions and arithmetic operations to avoid the time-consuming indexing operations in the bucketed relative attention bias, thereby accelerating the process. On the other hand, we found that the query-key attention map in self-attention might yield negative effects in recommendation tasks. Therefore, we removed the query-key attention map from the original self-attention layer and designed a new module called the Attention-free Token Mixer. Based on this framework, we propose a new model, FuXi-$\beta$. Comprehensive experiments demonstrate the remarkable efficiency of FuXi-$\beta$ and its 
outstanding performance. In the future, we plan to further reduce the complexity of the model and aim to address more complex recommendation tasks.

\bibliographystyle{ACM-Reference-Format}
\bibliography{ref}


\end{document}